\documentclass[pra,twocolumn,preprintnumbers,amsmath,amssymb,superscriptaddress]{revtex4-2}

\usepackage[utf8]{inputenc}
\usepackage[T1]{fontenc}
\usepackage{mathptmx}
\usepackage{etoolbox}
\usepackage{color}
\usepackage{hyperref}
\usepackage{ulem}
\usepackage{xcolor}
\usepackage{mathrsfs}
\usepackage{physics}

\PassOptionsToPackage{hyphens}{url} 
\usepackage{xurl}                   

\usepackage{microtype}              
\usepackage{hyperref}
\usepackage{soul}
\AtBeginDocument{\RenewCommandCopy\qty\SI}

\usepackage{relsize}
\usepackage{placeins}
\usepackage[T1]{fontenc}
\usepackage{amsmath}
\usepackage{array}
\usepackage{placeins}
\usepackage{siunitx}
\usepackage{booktabs}
\usepackage{rotating}
\usepackage[english]{babel}
\usepackage{csquotes}

\usepackage[utf8]{inputenc}
\usepackage[T1]{fontenc}
\usepackage{newtxtext,newtxmath}

\begin{document}


\title{Pre-emptive parametric kill switch for evaporative atomic sources in vacuum}

\author{Shuang Li}
\thanks{The indicated authors are joint first authors}

\affiliation{State Key Laboratory of Precision Spectroscopy, School of Physical and Material Sciences, East China Normal University (ECNU), Shanghai 200062, China}

\affiliation{New York University (NYU) Shanghai, 567 West Yangsi Road, Shanghai, 200126, China}

\author{Zhiyuan Lin}
\thanks{The indicated authors are joint first authors}

\affiliation{State Key Laboratory of Precision Spectroscopy, School of Physical and Material Sciences, East China Normal University (ECNU), Shanghai 200062, China}

\affiliation{New York University (NYU) Shanghai, 567 West Yangsi Road, Shanghai, 200126, China}

\author{Sen Li}
\affiliation{State Key Laboratory of Precision Spectroscopy, School of Physical and Material Sciences, East China Normal University (ECNU), Shanghai 200062, China}

\affiliation{New York University (NYU) Shanghai, 567 West Yangsi Road, Shanghai, 200126, China}

\author{Mohan Zhang}
\affiliation{State Key Laboratory of Precision Spectroscopy, School of Physical and Material Sciences, East China Normal University (ECNU), Shanghai 200062, China}

\affiliation{New York University (NYU) Shanghai, 567 West Yangsi Road, Shanghai, 200126, China}

\author{Fengquan Zhang}
\affiliation{State Key Laboratory of Precision Spectroscopy, School of Physical and Material Sciences, East China Normal University (ECNU), Shanghai 200062, China}

\affiliation{New York University (NYU) Shanghai, 567 West Yangsi Road, Shanghai, 200126, China}

\author{Jin Hu}
\affiliation{State Key Laboratory of Precision Spectroscopy, School of Physical and Material Sciences, East China Normal University (ECNU), Shanghai 200062, China}

\affiliation{New York University (NYU) Shanghai, 567 West Yangsi Road, Shanghai, 200126, China}

\author{Xiaotong Liu}
\affiliation{State Key Laboratory of Precision Spectroscopy, School of Physical and Material Sciences, East China Normal University (ECNU), Shanghai 200062, China}

\affiliation{New York University (NYU) Shanghai, 567 West Yangsi Road, Shanghai, 200126, China}

\author{Lin Meng}
\affiliation{State Key Laboratory of Precision Spectroscopy, School of Physical and Material Sciences, East China Normal University (ECNU), Shanghai 200062, China}

\affiliation{New York University (NYU) Shanghai, 567 West Yangsi Road, Shanghai, 200126, China}

\author{Tim Byrnes}
\email{tim.byrnes@nyu.edu}
\affiliation{State Key Laboratory of Precision Spectroscopy, School of Physical and Material Sciences, East China Normal University (ECNU), Shanghai 200062, China}

\affiliation{New York University (NYU) Shanghai, 567 West Yangsi Road, Shanghai, 200126, China}

\affiliation{\hbox{NYU-ECNU Institute of Physics at NYU Shanghai, 3663 North Zhongshan Road, Shanghai 200062, China}}

\affiliation{Center for Quantum and Topological Systems, New York University Abu Dhabi (NYUAD) Research Institute, NYUAD, UAE}

\affiliation{Department of Physics, New York University, New York, NY 10003, USA}

\author{Valentin Ivannikov}
\email{valentin@nyu.edu}
\affiliation{New York University (NYU) Shanghai, 567 West Yangsi Road, Shanghai, 200126, China}

\date{\today}

\begin{abstract}

A robust pre-emptive kill switch for cold atom experiments is introduced to significantly reduce costly system reassembly or replacement.
The design incorporates upper (alarm) and lower (evaporation) event detection mechanisms based on predefined thresholds.
Meanwhile, a duty cycle timing methodology is used to avert unintentional activation of the dispenser in circumstances where pulse signals occur.
The circuit employs generic components, a modular design, and formalized logic, ensuring cost-effectiveness, making the design suitable for school laboratories and other research environments.
This design is highly versatile and can be applied to other sensitive devices beyond dispensers, such as heating filaments, titanium sublimation pumps, tungsten lamps, and comparable systems.

\end{abstract}

\maketitle


\section{\label{sec:level1}Introduction}

Alkali metal vapors are a vital resource for numerous emerging chip-scale quantum technologies such as atomic clocks\nobreakspace\cite{filzinger2025ultralight, nichol2022elementary}, quantum gravimeters\nobreakspace\cite{cassens2025entanglement, szigeti2020high}, and quantum computers\nobreakspace\cite{ladd2010quantum,de2021materials}, which are increasingly attractive for the implementation of new quantum device architectures\nobreakspace\cite{wei2022collimated,martinez2023chip}. 
The sharp resonances of alkali atoms find applications in atom cooling, precision spectroscopy, and the frequency stabilization of lasers on atomic transitions\nobreakspace\cite{rusimova2019atomic, millerioux1994towards}. A source of (warm) alkali atoms is required in all cold-atom systems to provide an appropriate atom density to form the magneto-optic trap (MOT)\nobreakspace\cite{6matthews1998dynamical,7dugrain2014alkali,8barker2019single,ivannikov2018phase,ivannikov2013analysis}, as the appropriate atom density will significantly determine the characteristics of the MOT, such as load rate, loading time, loss rate, and cooling beam intensity. However, inappropriate density may even destroy MOT trapping\nobreakspace\cite{kang2018active}. 
Light-induced atom desorption (LIAD) from cell surfaces has been investigated to modulate atom density\nobreakspace\cite{christaller2022transient,19karaulanov2009controlling,20rusimova2019atomic,21hansel2001bose,22klempt2006ultraviolet,23gozzini1993light,24villalba2010light}. Although it performs well in some experiments, there can be reproducibility issues, 
and such methods have yet to be miniaturized to facilitate portability and microfabrication \cite{kang2017low,mcgilligan2020dynamic}.

There exists a wide range of methods to produce dilute atomic gases for 
alkaline, alkaline earth, rare earth, and other metals with low or moderate melting points \cite{rapol2001loading, slowe2005high,talker2021light, 13scherer2012characterization, sushkov2008production}. It is even possible to prepare vapors of more exotic metals that are particularly difficult to evaporate such as Zr, Nb, Mo, Ta and W by the dissociation of chemical compounds containing these metal atoms \nobreakspace\cite{1983}.
Commercial alkali-metal dispensers (e.g., Li, Na, K, Rb, Cs) have been widely used in laboratory-based cold-atom experiments for decades because they 
keep the background pressure low while providing many atoms for trapping\nobreakspace\cite{10kang2019magneto,11engler2000very}. Another advantage of using a metal dispenser over the more traditional atom sources in cold-atom experiments is that the dispenser can be used directly in an ultrahigh vacuum (UHV) environment\nobreakspace\cite{12ott2001bose}. 
Typically, the dispenser produces a controllable source of alkali-metal atom vapor emission from the dispenser into the UHV by evaporation during ohmic heating. The operation of the heater with a current in the range of 3–7 A heats the dispenser to a temperature range of 400-800°C\nobreakspace\cite{13scherer2012characterization,14succi1985atomic,16rapol2001loading}. 
We do not directly measure the dispenser temperature, even though it is tempting and we indeed have a thermocouple directly attached to the dispenser in situ, because temperature measurements are typically completed on the scale of the critical reaction time or slower, which makes such metrology inadequate for failure prevention. In contrary, the electrical current measurements are instant. If to limit the dispenser exposure to a capped current in time, it effectively limits the maximal energy allowed to be dissipated by the dispenser in a single run equal $T_{\text{max}} \cdot I^2 \cdot R_d$, where $I$ is the current, $R_d$ is the dispenser resistance, and $T_{\text{max}}$ is the longest allowed operation duration. The dissipated power is monotonically related to the temperature and this gives a way to exploit the instantaneous current measurement to judge about the temperature dynamic sufficient for regulatory purposes.

However, dispenser systems are characterized by complex nonlinear dynamics as a result of their open dissipative nature, making them vulnerable to instabilities that may induce hazardous conditions or equipment malfunctions. One such instability is exponential thermal runaway, which can result in dispenser burnout, a prevalent failure mechanism in titanium sublimation pumps and vacuum gauges, ultimately culminating in significant system failures. Therefore, the development of a robust kill switch for these sensitive devices is imperative 
to radically minimize the expensive system reassembly or replacement often causing devastating time losses.

While traditional Proportional-Integral-derived (PID) regulators are widely used for performing continuous-state control, they demonstrate inadequacies in handling abrupt transitions, such as power transistor failures, which can push the system into dangerous conditions. Moreover, improper tuning of PID parameters can exacerbate instability, thereby driving the system further from a safe state. As such, reliance solely on PID control is insufficient for the protection of these sensitive components. 
For example, similar mechanisms are already employed in modern battery systems, such as those in cell phones, to prevent overcharging and elongate battery life. These systems sometimes limit the charge to 80\% capacity before stopping and resuming the charge later. The prohibition of secondary activation under such conditions demonstrates an overcharge prevention approach that aligns conceptually with our design, albeit through simpler algorithms.
Furthermore, battery management systems use BNC (Battery Negative Current) and BMC (Battery Minus Current) modules to monitor and disconnect current through battery terminals if a threshold is exceeded, preventing overcurrent \cite{chen2022recent, bekhti2020design, cui2023thermal, xu2024series}. Therefore, a complementary protection mechanism is necessary to bridge this gap and ensure that the system transitions to a definitively safe state in all circumstances.

In this study, we develop a reliable kill switch utilizing the LTspice simulation, which was subsequently corroborated by building a physical prototype. The design ensures that the active state of the dispenser is safeguarded by implementing current limiters to ensure its operation within the prescribed current range. Synchronously, we devised a duty cycle timing strategy to avert unintended activation of the dispenser in response to pulse signal conditions.
The circuit functions as a quasi-DC safety mechanism, continuously supervising and discontinuing the electrical current should it surpass the established power dissipation parameters, characterized by the maximum current and specified operational duration constraints. This approach effectively limits the overall power dissipation, thus averting critical failures such as thermal runaway.

\section{Design overview}
\begin{figure}
\includegraphics{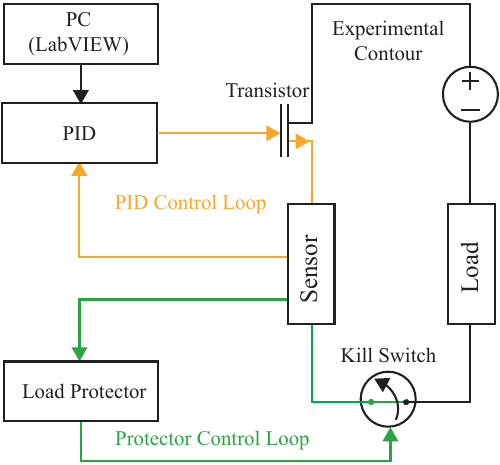}
\caption{\label{fig:flow chart}Generic experiment control}
\end{figure}

Our pre-emptive parametric kill switch design focuses on regulating the metal dispenser exposure through current limitation and ensuring precise activation timing.
As shown in Fig.\nobreakspace\ref{fig:flow chart}, the operational condition of the load is governed by two control loops: \textsc{PID Control Loop} and \textsc{Protector Control Loop}.
The first control loop is tasked with managing the current supplied to the transistor.  This loop cannot address potential discontinuous state transitions and fails to guarantee a safe state condition. In contrast, the secondary loop addresses these critical issues.

\begin{figure}
\includegraphics{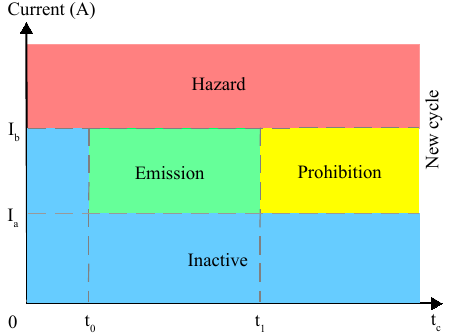}
\caption{\label{fig:phase}Phase diagram of the evaporation range of Rb-dispenser}
\end{figure}

As illustrated in Fig.\nobreakspace\ref{fig:phase}, the load current is maintained within the specified range $I_a - I_b$, and its activation time is accurately regulated by the duty cycle associated with the \textsc{Emission} and \textsc{Prohibition} periods. 
Following the \textsc{Emission} period, there exists a potential for the dispenser to be reactivated. The introduction of the \textsc{Prohibition} period eliminates this risk by preventing premature reactivation of the dispenser, thus maintaining the average dissipated energy within acceptable thresholds and ensuring the protection of the system.
Balancing heating and cooling requires tailoring the duty cycle to the specific dynamics of the experiment. The \textsc{Emission} period should be optimized during initial calibration to match the ideal heating and cooling time. This ensures fast cycling while maintaining dispenser safety. The \textsc{Prohibition}, activated sequentially after the \textsc{Emission}, should be optional via a switch. 
The likely pulse operation involves a single chamber that must be cleared of gasses before loading a new MOT.
In our setup, the dispenser cools faster than the cycle time because of heat conduction through massive copper wires. Note that this may not be applicable in other configurations.

Table~\ref{truth table} details the logical interconnections and operational parameters of the system shown in Fig.~\ref{fig:phase}. The experimental protocol begins with the \textsf{In1 Reset} signal, which functions to reset all system variables and states, thereby ensuring a clearly defined initial condition. Subsequently, the start of the state \textsc{Emission} is indicated by the \textsf{In2 EvapSt} signal at time \( t_0 \), during which the load current is maintained within a designated safety threshold.
It is to note that \( t_0 \) experiences a delay, which is ascribed to the thermal delay associated with heating the dispenser before the release of atoms. The \textsc{Emission} state represents the active period of atomic evaporation.
Upon the conclusion of this phase, the \textsf{In3 Prohib} signal terminates the \textsc{Emission} state and initiates the \textsc{Prohibition} state, thus enabling precise modulation of the duty cycle. 
Throughout the process, the \textsf{In4 Fail} signal continuously monitors the load current; should the current exceed the predefined threshold \( I_b \), the system transitions to the \textsc{Hazard} state as a protective measure.
The logical output exhibits a notable degree of logical precision. Specifically, the condition for the activation of the \textsf{Out1 Alarm} output is governed by the conjunction of the \textsf{In4 Fail} signal and the signal \textsf{In1 Reset} being false, as indicated by \( \overline{\textsf{Reset}} \). This interrelationship is formally articulated as:

\begin{equation}
\textsf{Alarm} = \textsf{Fail} \cdot \overline{\textsf{Reset}} .
\label{eq:alarm}
\end{equation}
The design logic substantially improves the reliability and safety of the system by ensuring the activation of the alarm exclusively upon the identification of a fault condition (\textsf{Fail}), contingent upon the system not being in a reset state (\(\overline{\textsf{Reset}}\)). The alarm remains active until manually deactivated, thus avoiding unnoticed fault conditions and mitigating the risk of undetected system failure during operation.

The activation of \textsf{Out2 Evap} is dependent upon more rigorous conditions, which involve a logical integration of four input signals. The corresponding expression is formulated as follows:
\begin{equation}
\textsf{Evap} = \overline{\textsf{Fail}} \cdot \overline{\textsf{Reset}} \cdot \textsf{EvapSt}\left(t = -\Delta t_1\right) \cdot \overline{\textsf{Prohib}}\left(t = -\Delta t_2\right) ,
\label{eq:evap}
\end{equation}
where \textsf{Evap} signifies the initiation of the evaporation process. The \textsf{EvapSt} signal delineates the beginning of the zone \textsc{Emission} as depicted in Fig.~\ref{fig:phase}, where its evaluation at \( t = -\Delta t_1 \) determines its duration. Analogously, \textsf{Prohib} marks the beginning of the \textsc{Prohibition} zone, with the term \( t = -\Delta t_2 \) determining its length. Therefore, the activation of the evaporation process is contingent upon: (1) the absence of any failure or reset, (2) the prior initiation of the evaporation process at \(\Delta t_1\), and (3) the non-activation of the prohibition condition for \(\Delta t_2\). This logical framework enforces strict conditions for initiating the evaporation process, ensuring compliance with safety and operational constraints.

\begin{figure}[htbp]
\centering
\includegraphics{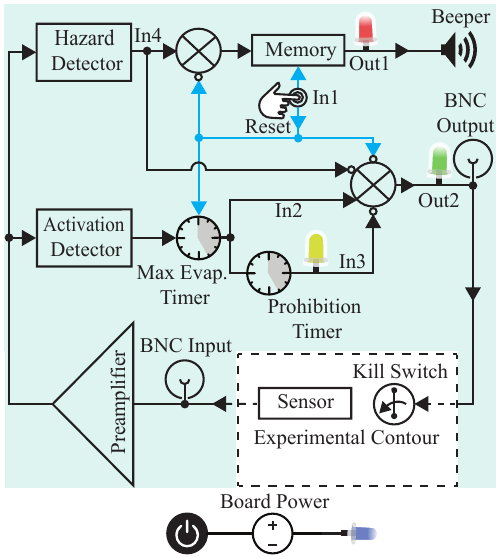}
\caption{\label{fig:Functional solution} Functional solution diagram. The light green background area corresponds to the \textsc{Load Protector} in Fig.~\ref{fig:flow chart}. LEDs are color-coded identically to the phase diagram of Fig.~\ref{fig:phase}.}
\end{figure}

Drawing upon the logical framework of Table~\ref{truth table} and the dynamic time-current characteristics of Fig.~\ref{fig:phase},  we show a functional representation of the operational behavior of the system in Fig.~\ref{fig:Functional solution}. This diagram provides the foundation for the practical realization of the system.
The next section elaborates on the design and implementation of a circuit that executes the designated logical functions. We highlight the importance of selecting appropriate logic gates and optimizing their arrangement to ensure reliable performance and resource efficiency.

\begin{table}[htb]
    \centering
    \caption{\label{truth table}Rb-dispenser protection circuit truth table.}
    \begin{ruledtabular}
    \begin{tabular}{c|cccc|cc}
         
         \textbf{Scenario} & \textbf{In4} & \textbf{In3} & \textbf{In2} & \textbf{In1} & \textbf{Out1} & \textbf{Out2} \\
         
        \textbf{\textnumero} & \textbf{Fail} & \textbf{Prohib} & \textbf{EvapSt} & \textbf{Reset} & \textbf{Alarm} & \textbf{Evap} \\
        \hline
        1  & 0 & 0 & 0 & 0 & 0 & 0 \\
        2  & 0 & 0 & 0 & 1 & 0 & 0 \\
        3  & 0 & 0 & 1 & 0 & 0 & 1 \\
        4  & 0 & 0 & 1 & 1 & 0 & 0 \\
        5  & 0 & 1 & 0 & 0 & 0 & 0 \\
        6  & 0 & 1 & 0 & 1 & 0 & 0 \\
        7  & 0 & 1 & 1 & 0 & 0 & 0 \\
        8  & 0 & 1 & 1 & 1 & 0 & 0 \\
        9  & 1 & 0 & 0 & 0 & 1 & 0 \\
        10 & 1 & 0 & 0 & 1 & 0 & 0 \\
        11 & 1 & 0 & 1 & 0 & 1 & 0 \\
        12 & 1 & 0 & 1 & 1 & 0 & 0 \\
        13 & 1 & 1 & 0 & 0 & 1 & 0 \\
        14 & 1 & 1 & 0 & 1 & 0 & 0 \\
        15 & 1 & 1 & 1 & 0 & 1 & 0 \\
        16 & 1 & 1 & 1 & 1 & 0 & 0 \\
    \end{tabular}
    \end{ruledtabular}
\end{table}

\begin{figure*}[htbp]
\centering
\includegraphics[width=\textwidth]{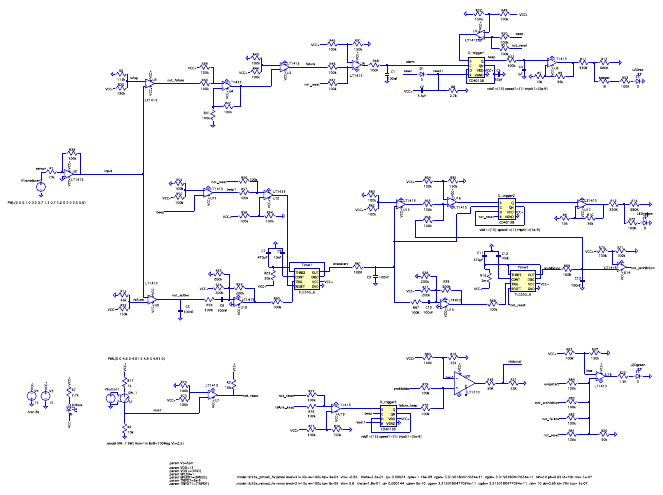}
\caption{\label{fig:dispenser_protection}The design circuit of the pre-emptive parametric kill switch.}
\end{figure*}

\begin{figure}[htbp]
\centering
\includegraphics[width=\columnwidth]{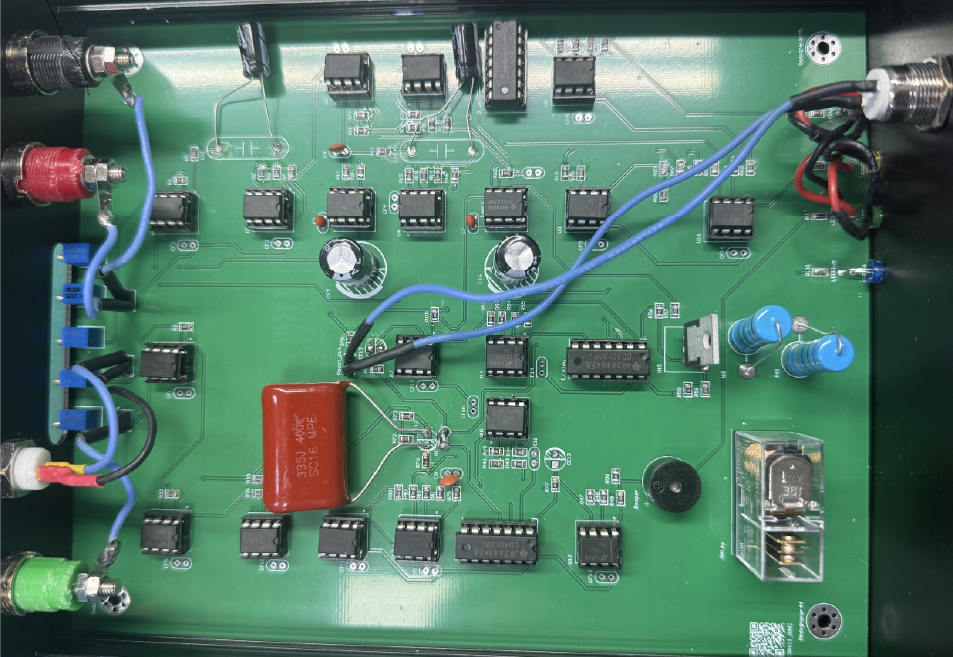}
\caption{\label{fig:PCB} PCB with dimensions of 192~mm $\times$ 154~mm, designed and verified in LTspice.}
\end{figure}

\begin{figure}[htbp]
\centering
\includegraphics[width=\columnwidth]{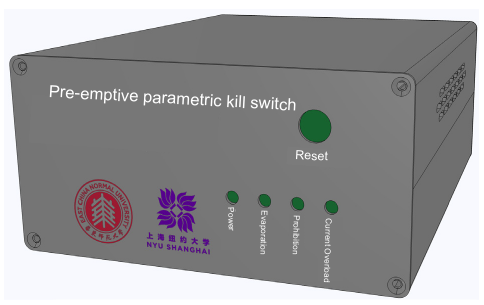}
\caption{\label{fig:Aluminum alloy enclosure} Aluminum alloy enclosure (1.5 mm thickness, dimensions: 160 mm (L) × 192.6 mm (W) × 80 mm (H)) designed to house the PCB shown in Fig.~\ref{fig:PCB}, the front panel includes openings for four LED indicator lights and a reset button.}
\end{figure}

\begin{figure}[htbp]
\centering
\includegraphics[width=\columnwidth]{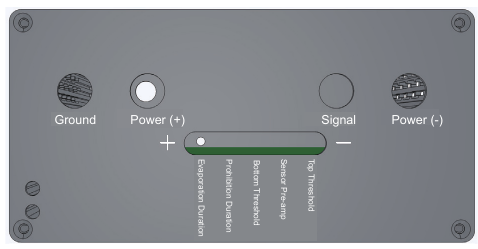}
\caption{\label{fig:Rear cover} Rear cover of the aluminum alloy enclosure in Fig.~\ref{fig:Aluminum alloy enclosure} , featuring power input terminals, a ground port, and potentiometer adjustment holes.}
\end{figure}

\section{circuit design}

Fig.~\ref{fig:PCB} shows the PCB (192~mm $\times$ 154~mm ) designed with LTspice, housed in the aluminum enclosure shown in Fig.~\ref{fig:Aluminum alloy enclosure}. 
The front panel has apertures for four LEDs and a reset function, the rear panel contains power terminals, a ground port, and five adjustment holes for the calibration of the PCB potentiometer, as seen in Fig.~\ref{fig:Rear cover}.  The circuit bill of materials is listed in Table~\ref{budget}.
The detailed circuit schematic is shown in Fig.~\ref{fig:dispenser_protection}. 
The blue LED lights up, which indicates that the PCB board is powered on. 
The \textsf{Reset} signal originates from PWL (Piecewise Linear) high-level pulses controlled by LabVIEW. The \textsf{SW\_1} switch is used for manual reset. \textsf{U1} inverts the \textsf{Reset} signal to create the \textsf{not\_reset} signal, serving as an active low-reset for subsequent circuits.
The sensor signal is transmitted to the preamplifier \textsf{U2} through the potentiometer \textsf{R1}, which functions to optimize the operational range of the sensor. Upon integration of an \textsc{input}, the circuit functions in two distinct modes as described below, where this parallel configuration design effectively prevents any potential competition or interference among signals.

In one of the operational modes, the circuit is designed to look for when the current goes beyond an upper limit \( I_a \). Once that point is reached, the protection path begins to work to keep the dispenser from drawing more current than the safety margin allows. The comparator \textsf{U3} handles this job, with its reference level adjusted through the potentiometer \textsf{R6}. The output of \textsf{U3}, named \textsf{not\_failure}, is then shaped by \textsf{U4} together with the resistors \textsf{R39--R42}, which help to clean up the transitions and make the signal less sensitive to noise. The next stage, \textsf{U5}, simply flips the logic and produces what we call the active \textsf{failure} signal.
After that, \textsf{U6} works as an AND gate, combining both \textsf{failure} and \textsf{not\_reset}. The signal that comes out passes through capacitor \textsf{C1}, which smooths it and eliminates bounce or short spikes, giving the final \textsf{Alarm} output as shown in Eq.\nobreakspace\eqref{eq:alarm}. The \textsf{Alarm} acts as the clock input for \textsf{D\_trigger1}. Meanwhile, the external \textsf{reset} line goes through the Zener diode \textsf{D1}, forming the internal node \textsf{reset1}. This node connects to the reset terminal of \textsf{D\_trigger1} and to the RC branch made of \textsf{C2} and \textsf{R8}, which performs the power reset. The diode \textsf{D1} isolates the RC path so that any stored charge does not leak back, and the external reset still works normally.
The output from \textsf{D\_trigger1}, named \textsf{beep}, goes through \textsf{U8}, which provides enough current to drive the beeper and the red LED indicator (\textsf{LEDred}).
Another part of the circuit deals with the \textsf{beep} signal. Here, \textsf{U7} takes the logical combination of \textsf{beep} and \textsf{not\_reset} to make the input of the set for \textsf{D\_trigger1}, following \(
\textsf{S} = \textsf{beep} \cdot \overline{\textsf{Reset}}.
\label{S}
\)
Consequently, this arrangement ensures that when the input voltage crosses the set limit, the alarm remains on until someone manually resets the system. In this way, the circuit does not recover itself when a fault occurs.

The other operational mode involves detecting when the dispenser reaches a predetermined critical minimum threshold \( I_b \) and avoiding the risk that the dispenser is activated for prolonged periods due to malfunctions or unforeseen current surges.
The circuit incorporates a comparator \textsf{U9} to detect the crossing of a predefined lower threshold \( I_b \) (the evaporation threshold). The reference voltage for \textsf{U9} is established using the potentiometer \textsf{R15}.
The \textsf{U9} output signal is filtered by an RC circuit (\textsf{R59}, \textsf{C3}) to suppress noise and eliminate signal bounce, then compared to a 0V reference by the comparator \textsf{U10} to convert it into a stable digital logic level signal. 
This signal is fed into the \textsf{Timer1} module to set the duration ($ \Delta T_1$) of activation of the dispenser, where \( \Delta T_1 \approx 1.1 \times C4 \times R23 \), 
\textsf{R23} functions as a potentiometer that offers variable resistance to modulate the time constant.
Module \textsf{U11} modifies the \textsf{beep} signal to create \textsf{beep1}, and \textsf{U12} integrates \textsf{beep1} with the \textsf{not\_reset} signal to generate a stable control level that drives the RESET pin of \textsf{Timer1}, deciding whether the timer operates or remains reset.
The \textsf{Timer1} output is designated as \textsf{evapstart}.
To avoid the risk of unintended activation of the dispenser due to transient pulse currents, the \textsf{evapstart} signal splits into two paths:
the first path processed by \textsf{U13} (configured with identical components of the RC network as \textsf{U10}: \textsf{R27--R29}, \textsf{R67} and \textsf{C7}) feeding \textsf{Timer2} which sets the duration \( \Delta T_2 \approx 1.1 \times C8 \times R16 \), where \textsf{R16} is implemented as a potentiometer. 
Its output designated as \textsf{prohibition} is inverted and shaped by \textsf{U14} with peripheral resistors \textsf{R70} and \textsf{R71} to generate a stable \textsf{NOT-prohibition} signal. While the second path routes \textsf{evapstart} through \textsf{U15} for signal conditioning, then integrated with \textsf{prohibition} in \textsf{U16} before transmission to the \textsf{D\_trigger2}, which drives \textsf{U17} as a buffer amplifier to illuminate the \textsf{LEDyellow} indicating the activation status of \textsf{Timer2}. In short, the configuration helps prevent the risk of prolonged activation that could affect the performance and service life of the dispenser.

The ultimate configuration of the circuit includes a quad-input logic AND gate \textsf{U18}, which integrates the control signals \textsf{evapstart}, \textsf{not\_prohibition}, \textsf{not\_failure}, and \textsf{not\_reset} to activate the \textsf{LEDgreen} and the \textsf{relay}. The mathematical rationale underlying the operation of the gate is elucidated in Eq.\nobreakspace\eqref{eq:evap}. This approach eliminates the need for extra driver chips, reducing components and board complexity while maintaining full functionality.
For gradual development convenience, we kept one spare (unused) opamp in each dual-opamp chip.

\begin{table*}[htb]
    \centering
    \caption{\label{budget}Bill of materials}
    \scalebox{0.65}{
    \begin{tabular}{*{9}{c}}
        \hline\hline
        \textbf{\textnumero} & \textbf{Part name} & \textbf{Manufacturer} & \textbf{Model ID} & \textbf{Tag} & \textbf{Characteristics} & \textbf{Packaging/Technology} & \textbf{Qty} & \textbf{USD/unit} \\
        \hline
        1 & Relay & Omron & G2R-1 DC5 & - & 5 V(DC), 10 A & SPDT & 1 & 3.430 \\
        
        2 & D-trigger & Texas Instruments & CD4013BE & Dtrigger & 45ns, 12MHz, 3V$-$18V, 6.8 mA & PDIP-14 & 2 & 0.530 \\
        
        3 & Timer & Texas Instruments & TLC555 & Timer1, Timer2 & 5V$-$15V & PDIP-8 & 2 & 0.300 \\
        
        4 & Operational Amplifier & Analog Devices & LT1413CN8\#PBF & U1-19 & 950kHz, 0.4 V/\textmu s, $\pm 2$V$-\pm 18$V & PDIP-8, 2 channels & 17 & 6.420 \\
        
        5 & Operational Amplifier & Analog Devices & LT1210CT7\#PBF & U20 & 8V-36V, $\pm 5$V-$\pm 15$V , 350 mA & TO-220, 1 channel & 1 & 7.950 \\
        6 & Beeper & Multicomp Pro & ABT-414-RC & beeper & $3-8$V, 40mA, 85dB & - & 1 & 1.050 \\
        
        
        7 & LED & Broadcom & HLMP-3301 & LEDRed & Red, $2.0-2.2$V,20mA & - & 1 & 0.128 \\
        
        8 & LED & Broadcom & HLMP-3401 & LEDYellow & Yellow, $2.0-2.2$V,20mA & - & 1 & 0.128 \\
        
        9 & LED & Broadcom & HLMP-3507 & LEDGreen & Green, $2.0-2.2$V,20mA & - & 1 & 0.128 \\
        
        10 & LED & Broadcom & HLMP-LB65-RU0DD & LEDBlue & Blue, $3.0-3.2$V,20mA & - & 1 & 0.497 \\
        
        11 & Resistor & TE Connectivity  & 1622829-1 & R2-5 & $10$k$\Omega$, 50V,0.1 W & Surface Mount & 4 & 0.007 \\

        12 & Resistor & TE Connectivity & CRGCQ0402F2K7 & R7,8 & $2.7$k$\Omega$, 0.063 W & Surface Mount & 2 & 0.007 \\

        13 & Resistor & TE Connectivity & CRG0402F36K & R9,10 & $36$k$\Omega$, 0.063W & Surface Mount & 2 & 0.007 \\

        14 & Resistor & TE Connectivity & CRG0402F330K & R11,12 & $330$k$\Omega$, 0.063W & Surface Mount & 2 & 0.007 \\

        15 & Resistor & TE Connectivity & CRG0402F680K & R13,14 & $680$k$\Omega$, 0.063W & Surface Mount & 2 & 0.004 \\
        
        16 & Resistor & TE Connectivity & 1-1622826-8 & R17 & $1$k$\Omega$, 0.5W & Surface Mount & 1 & 0.006 \\

        17 & Resistor & TE Connectivity & CRGH2512J22K & R18 & $22$k$\Omega$, 2 W, 200V & Surface Mount & 1 & 0.013 \\

        18 & Resistor & Vishay & PNM0805E5002BST5 & R19,20 & $50$k$\Omega$, 0.2W & Surface Mount & 2 & 0.446 \\

        19 & Resistor & TE Connectivity & AMP 1614911-6 & R21 & $25$k$\Omega$, 0.1W & Surface Mount & 1 & 0.129 \\

        20 & Resistor & TE Connectivity & CRGCQ0402F1K5 & R22 & $1.5$k$\Omega$, 0.063W & Surface Mount & 1 & 0.006 \\

        21 & Resistor & TE Connectivity & CRG0402F200K & R24-29 & $200$k$\Omega$, 0.063W & Surface Mount & 4 & 0.003 \\
         
        22 & Resistor & TE Connectivity & CRG0402F150K & R30-37 & $150$k$\Omega$, 0.063W & Surface Mount & 5 & 0.006 \\

        23 & Resistor & TE Connectivity & CRG0402F100K & R38-85, & $100$k$\Omega$, 0.063W & Surface Mount & 11 & 0.007 \\
        
        24 & Cermet trimmer potentiometer & Bournes & 3296W-1-103LF & R1 & $10$k$\Omega,0.5W$ & 25 turns, Cermet,THT & 1 & 1.238 \\
        
        25 & Cermet trimmer potentiometer & Bournes & 3296W-1-204LF & R6 & $200$k$\Omega$,0.5W & 25 turns, Cermet,THT & 1 & 1.192 \\

        26 & Cermet trimmer potentiometer & Bournes & 3296W-1-503LF & R15 & $50$k$\Omega$,0.5W & 25 turns, Cermet,THT & 1 & 1.145 \\

        27 & Cermet trimmer potentiometer & Bournes & 3296W-1-205LF & R16 & $2$m$\Omega$,0.5W & 25 turns,  Cermet,THT & 1 & 1.190 \\
          
        28 & Cermet trimmer potentiometer & Bournes & 3299Y-1-203LF & R23 & $20$k$\Omega$,0.5W & 25 turns, Cermet,THT & 1 & 1.470 \\

        29 & Silicon Rectifier Diode & NTE Electronics & NTE507 & D1 & 1A, 50V & Through Hole & 1 & 0.675 \\

        30 & Multilayer Ceramic Capacitor & Vishay & K103K15X7RH53L2 & C5,C9 & 10 nF, 100V & Radial Leaded & 2 & 0.079 \\

        31 & Plastic Film Capacitor & Panasonic & ECWFD2W335 & C2 & $3.3 \mu$F, 400V &  Radial Leaded & 1 & 0.399 \\
         
        32 & Multilayer Ceramic Capacitor & Vishay & K104K20X7RH5TL2 & C1,C6 & 100nF, 100V &  Radial Leaded & 2 & 0.137 \\
        
        33 & Aluminum Electrolytic Capacitor & Panasonic & EEUTA1C471 & C4,C8 & $470\mu$F & Radial Leaded & 2 & 0.484 \\
        
        34 & Aluminum Electrolytic Capacitor & Panasonic  & ECA2DHG470 & C3,C7 & $47\mu$F & Radial Leaded & 2 & 0.024 \\
        
        35 & DIP Socket & TE Connectivity  & 1-2199298-2 & - & 1~A,250~V & 8P & 19 & 0.103 \\

        36 & DIP Socket & SparkFun Electronics  & PRT-07939 & - & - & 14P & 3 & 0.600 \\
        
        37 & Banana Jack & Cinch Connectivity Solutions & 108-0903-001 & - & 15 A, 7 kVDC  & Female & 3 & 0.710 \\

        38 & Pushbutton & Adam Tech & SW-PB1-1DZ-A-P1-A & SW\_1 & 3A,250V & solder type & 1 & 0.235 \\
        
        39 & PCB & Phoenix Contact  & UM-BASIC 108/32 DEV-PCB & - & 10cm $\times$ 7cm & - & 1 & 0.505 \\

        40 & PCB-mounting screws & - & - & - & L=$25$mm,D=$3.2$mm & - & 4 & 0.150 \\

        41 & Aluminum alloy enclosure & - & - & - & $160\,\mathrm{mm} \times 192.6\,\mathrm{mm} \times 80\,\mathrm{mm}$ & 1.5 mm thickness & 1 & 18.860 \\
        \hline
        \multicolumn{8}{r}{\textbf{Total}} & \textbf{160.18} \\
        \hline\hline
    \end{tabular}
    }
\end{table*}

\section{SUMMARY AND CONCLUSIONS}

In this work, we developed a protection circuit that proved reliable in real-world operation, particularly for cold atom experiments. 
The primary purpose was to maintain the dispenser current within a safe region (\(I_a - I_b\)) and to prevent accidental triggering caused by noise or sudden pulses. 
To do that, the circuit uses a duty–cycle timing control: it only allows emission during \( \Delta T_1 \) and then automatically cuts off the current in the following period \( \Delta T_2 \). 
This small change makes the whole setup much safer by preventing overheating and extending the life of the dispenser.
Instead of using digital logic or a controller, we kept the whole design analog. 
While this may seem ``old-fashioned'', it eliminates the potential for failure due to complex protection in cases where computer or controller implementations are prone to hanging or glitching due to signal racing. 
All parts of the circuit are standard and inexpensive, so it is easy to reproduce or repair. 
This makes it very practical for university labs or any small experimental setup.
Our design may be applied to many other devices that share similar thermal or current-safety concerns—such as heater filaments, titanium sublimation pumps, or tungsten lamps. 
With a few adjustments, this protection logic could even scale up to larger applications, such as battery modules or high-power reactors, where shutting the system down early can prevent far more serious failures.


\FloatBarrier

\begin{acknowledgments}

This work is supported by the SMEC Scientific Research Innovation Project (2023ZKZD55); the Science and Technology Commission of Shanghai Municipality (22ZR1444600); the NYU Shanghai Boost Fund; the China Foreign Experts Program (G2021013002L); the NYU-ECNU Institute of Physics at NYU Shanghai; the NYU Shanghai Major-Grants Seed Fund; and Tamkeen under the NYU Abu Dhabi Research Institute grant CG008.

\end{acknowledgments}

\section *{data availability}

LTspice files for designing the circuit board can be obtained from 
\href{https://github.com/Shuang1215/KillSwitch}
     {GitHub repository (\nolinkurl{https://github.com/Shuang1215/KillSwitch})}

\nocite{*}

\bibliographystyle{apsrev}
\bibliography{ref}

\end{document}